\documentclass[11pt]{article}
\pdfoutput=1
\usepackage{jcappub,natbib}
\def\la{\mathrel{\raise.3ex\hbox{$<$\kern-.75em\lower1ex\hbox{$\sim$}}}}

\title{Instability in axion inflation with strong backreaction from gauge modes} 

\author[a,b]{Marco Peloso}
\author[c]{Lorenzo Sorbo}

\affiliation[a]{Dipartimento di Fisica e Astronomia ``Galileo Galilei'', Universit\`a di Padova, 35131 Padova, Italy}
\affiliation[b]{INFN, Sezione di Padova, 35131 Padova, Italy}
\affiliation[c]{Amherst Center for Fundamental Interactions, Department of Physics,
University of Massachusetts, Amherst, MA 01003, U.S.A.}

\abstract{We perform an analytical study of the stability of the background solution~\cite{Anber:2009ua} of the model in which an inflaton, through an axionic coupling to a $U(1)$ gauge field, causes an amplification of the gauge field modes that strongly backreact on its dynamics. To this goal, we study the evolution of the gauge field modes coupled to the inflaton zero mode, treating perturbatively the deviation of the inflaton velocity from its mean-field value. As long as the system is in the strong backreaction regime we find that the inflaton velocity performs oscillations of increasing amplitude about the value it would have in the approximation of constant velocity, confirming an instability that has been observed in numerical studies.
}

\begin{document}

\maketitle
\flushbottom

\section{Introduction}
\label{sec:intro} 

Primordial inflation is the leading candidate for the description of the earliest moments of the Universe which are currently accessible to observation. The fundamental ingredient of this framework is a period of accelerated expansion, typically fueled by a scalar field, the inflaton, that is slowly rolling down its potential $V(\Phi)$. While the simplest way of modeling inflation consists simply of choosing the shape of $V(\Phi)$, a more realistic description should also account for the interactions of the inflaton with other degrees of freedom. These degrees of freedom can be excited by the time-dependence of the rolling inflaton. As a consequence of energy conservation, such a process of particle production will generally slow down the rolling of the inflaton. 

While the excited modes can have perturbatively small effects, in some models and/or for some choices of parameters their backreaction can significantly affect the evolution of the inflaton. The slowing down of the inflaton can be so effective as to allow inflation even if the inflaton potential does not satisfy the slow-roll conditions $|V'|\ll V/M_P$, $V''\ll V/M_P^2$. This can be very helpful for inflationary model building, as for instance finding corners of the String Theory Landscape where the slow-roll conditions are satisfied has proven very difficult.

These mechanisms can be seen at work in models of warm inflation~\cite{Berera:1995ie} where the particles produced during inflation are assumed to follow a thermal distribution. The model of trapped inflation~\cite{Green:2009ds} provides a scenario where, on the other hand, the produced particles do not thermalize, but the microphysical mechanism underlying particle production is spelled out in detail.

In this paper we focus on the model of natural steep inflation of Anber and Sorbo (AS)~\cite{Anber:2009ua}, which has the advantages of having a well defined microphysical description and of being described by relatively simple formulae. In this model, which we review in Section~\ref{sec:background} below, a pseudoscalar inflaton is coupled to a $U(1)$ gauge field via a dimension-$5$ operator $\frac{\alpha}{4f}\Phi F_{\mu\nu}\tilde{F}^{\mu\nu}$, where $\alpha/f$ is a constant. As a consequence of this coupling, the modes of one of the helicities of the gauge field get amplified by a factor $\propto \exp\{\frac{\pi}{2}\frac{\alpha\dot{\bar\Phi}}{fH}\}$, where $H$ is the Hubble parameter during inflation and $\dot{\bar\Phi}$ denotes the velocity of the zero mode of the inflaton. 

Due to the exponential dependence of its amplitude on $\dot{\bar\Phi}$, the gauge field can strongly backreact on the zero mode of the rolling inflaton. This effect is due to a term  $\frac{\alpha}{4f} F_{\mu\nu}\tilde{F}^{\mu\nu}$ on the right-hand side of the Klein-Gordon equation that controls the evolution of the zero mode of the inflaton. In~\cite{Anber:2009ua} the effect of this term was evaluated in the approximation of constant $\dot{\bar\Phi}$, which leads to a slow-roll solution $\dot{\bar\Phi}=\frac{f\,H}{\alpha}$ times a factor, of order $10\div 100$, that has a logarithmic dependence on the parameters in the theory. This result leads to conclude that slow roll inflation can be obtained in this model even if the potential is steep and does not satisfy the slow-roll conditions. 

The amplitude of the gauge field modes has an exponential dependence on $\dot{\bar\Phi}/H$, which is a quantity that generally increases during inflation. As a consequence, even if the backreaction of the gauge modes is negligible at early stages of inflation (e.g., when observable CMB scales leave the horizon), it will generally become important towards the end of inflation, at a time when gravitational waves \cite{Cook:2011hg,Barnaby:2011qe,Domcke:2016bkh} or a population of primordial black holes \cite{Linde:2012bt,Bugaev:2013fya,Garcia-Bellido:2016dkw,Garcia-Bellido:2017aan} might be generated.  For these reasons it is important to have a reliable description of the behavior of the system in the strong backreaction regime.

Several works~\cite{Cheng:2015oqa,Notari:2016npn,DallAgata:2019yrr,Domcke:2020zez,Caravano:2022epk,Gorbar:2021rlt}  have studied the behavior of this inflaton/gauge field  system in the strong backreaction regime, using numerical techniques beyond the constant $\dot{\bar\Phi}$ approximation but still, with the exception of~\cite{Caravano:2022epk}, neglecting the spatial dependence of the inflaton\footnote{Additional lattice studies of the behavior of this inflaton-gauge field system at the end of inflation can be found in~\cite{Adshead:2015pva,Adshead:2016iae,Figueroa:2017qmv,Adshead:2018doq,Cuissa:2018oiw,Adshead:2019lbr,Adshead:2019igv,Figueroa:2021yhd}.}. These studies do consistently show, in the strong backreaction regime, large and apparently irregular oscillations in the velocity of the inflaton, in contrast to the smooth, monotonic behavior found in the analytical studies where $\dot{\bar\Phi}$ is treated as a constant. While most of these studies reported the existence of these oscillations without explaining their origin, in reference~\cite{Domcke:2020zez} this behavior has been linked to the fact that the backreaction term gives a delayed response to the changes of the velocity of the inflaton.

In this work we explore the behavior of the model of~\cite{Anber:2009ua} in the strong backreaction regime, going beyond the approximation used in~\cite{Anber:2009ua}, but without resorting to numerical techniques.  This can allow to better understand the parameter dependence of the model. Moreover, many of the existing numerical solutions start from a (stable) regime of small backreaction, which then evolves into the strong backreaction AS regime, so one could still imagine that the steady-state AS background solution is indeed stable, but with a very narrow basin of attraction which is not probed by the existing numerical solutions. For these reasons, we perform an analytic study of the linearized stability of the AS background solution, starting from field configurations that deviate from it only by a perturbatively small amount. As in most of the numerical investigations, we also neglect the spatial dependence of the inflaton. On the one hand, the qualitative agreement between~\cite{Cheng:2015oqa,Notari:2016npn,DallAgata:2019yrr,Domcke:2020zez,Gorbar:2021rlt} and~\cite{Caravano:2022epk} leads to believe that this ingredient is not crucial for the claimed instability of the AS solution. On the other hand, this assumption greatly simplifies our computation, making it  more transparent. 

Our analysis shows that in the regime of strong backreaction the evolution of the zero-mode of the inflaton is indeed unstable, and that oscillations of increasing amplitude develop. Moreover, it shows, consistently with the analysis of~\cite{Domcke:2020zez}, that the backreaction term does indeed respond with a delay to the changes in $\dot{\bar\Phi}$, which can be seen as the cause of the oscillations.

The paper is structured as follows. In Section \ref{sec:background} we review the AS background solution. In Section \ref{sec:perturbations} we compute analytically (with some approximations) the evolution of linear perturbations about this solution. The computation is divided in three parts. We first introduce the perturbations and the linearized system of equation that describe their evolution. 
Then, in Subsection \ref{subsec:green} we obtain the Green function to express the gauge fluctuations as a functional of the inflaton fluctuation. Then in Subsection \ref{subsec:source+sol} we provide the approximate solution for the latter. In Section \ref{sec:conclusions} we present our conclusions. This section is followed by two appendices. In Appendix \ref{app:WKB} we derive the AS gauge modes with a WKB approximation. In Appendix \ref{app:source} we evaluate the source term describing the backreaction of the gauge fluctuations to the inflaton fluctuation. 

\section{Background solution} 
\label{sec:background}

The mechanism of~\cite{Anber:2009ua}  is characterized by the action 
\begin{equation} 
S = \int d^4 x \sqrt{-g} \left[ - \frac{1}{2} \left( \partial \Phi \right)^2 - V \left( \Phi \right) 
- \frac{1}{4} F^2 - \frac{\alpha}{4 f} \Phi F {\tilde F} \right] \;, 
\end{equation}
where $\Phi$ is the axion inflaton, with potential energy $V$, while $F$ is the field strength of a $U(1)$ field and ${\tilde F}$ is its dual. The quantity $f$ is the (mass dimension one) axion gauge constant, while $\alpha$ is the dimensionless parameter controlling the coupling of $\Phi$ to the specific gauge field that we are considering. We use conformal time and we neglect slow roll corrections to the background geometry 
\begin{equation}
g_{\mu \nu} = a^2 \left( \tau \right) \, {\rm diag } \left( -1 ,\, 1 ,\, 1 ,\, 1 \right) \;\;,\;\; a = - \frac{1}{H \tau} \,, 
\end{equation}
where $H$ is the inflationary Hubble rate. In the background solution of  \cite{Anber:2009ua}  the inflaton is taken to be homogeneous, while the gauge fields has both time and spatial dependence. In the $A_0 = \vec{\nabla}\cdot\vec{A}=0$ gauge we decompose 
\begin{equation}
\vec{A} \left( \tau ,\vec{x} \right)= \sum_{\lambda = \pm} \int \frac{d^3 k}{\left( 2 \pi \right)^3} \left[ \vec{\epsilon}_\lambda \left( \vec{k} \right) A_\lambda \left( \tau ,\, k \right) {\hat a}_\lambda \left( \vec{k} \right) {\rm e}^{i \vec{k} \cdot \vec{x}} + {\rm h.c.} \right] \,, 
\end{equation}
where $\vec{k}$ is the comoving momentum, ${\hat a}_\pm \left( \vec{k} \right)$ annihilation operators, and $\vec{\epsilon}_\pm$ are circular helicity operators, obeying $\vec{k} \cdot \vec{\epsilon}_\pm = 0$ and $\vec{k} \times \vec{\epsilon}_\pm = \mp i \, k \, \vec{\epsilon}_{\pm}$. 

Denoting by a prime the derivative with respect to conformal time, the homogeneous inflaton and the vector field obey the equations~\cite{Anber:2009ua} 
\begin{eqnarray}
&& \Phi'' + 2 a H \Phi' + a^2 \, V' = - \frac{\alpha^2}{4 \pi^2 a^3 \, f} \int dk \, k^2 \, \frac{\partial}{\partial \tau} \left[\left\vert A_+ \right\vert^2-\left\vert A_- \right\vert^2\right] \;, \nonumber\\ 
&& A_\lambda '' + k^2 A_\lambda - \frac{\lambda \, \alpha \, \Phi'}{f} \, A_\lambda = 0 \;, 
\label{master}
\end{eqnarray} 
where prime on the inflaton potential denotes differentiation with respect to $\Phi$. 

The goal of this work is to solve this system of equations for the background solution of \cite{Anber:2009ua}  and for small departures about it. We first solve the equation for the gauge field, that thus becomes a functional of $\Phi'$. We then insert the formal solution into the first equation, that this way  becomes an integro-differential equation for the homogeneous inflaton. In this section we review the solution of \cite{Anber:2009ua} for which we denote 
\begin{equation}
A_\lambda = {\bar A}_\lambda  \;\;,\;\; 
\Phi = {\bar \Phi} \;\;,\;\; {\rm with} \;\; \xi \equiv \frac{\alpha \, {\bar \Phi'}}{2 f a H} = {\rm const} \;. 
\end{equation} 

Depending on the sign of ${\bar \Phi}'$, one of the two helicity modes experiences an unstable growth next to horizon crossing \cite{Anber:2009ua}. For definiteness (and without loss of generality), we assume that ${\bar \Phi}' > 0$ so that the unstable mode is the $\lambda = +$ one. This implies that $V' < 0$. We disregard the stable $A_-$ mode and, to shorten notation, from now on we relabel $A_+ \equiv A$. The equation for the mode functions then admits the two linearly independent solutions that, in the long wavelength limit, can be written as
\begin{eqnarray}
&& {\bar A}_1 \left( \tau  ,\, k \right) \Big\vert_{x \ll 2 \xi} \simeq \frac{1}{\sqrt{2 k}} \left[ 
\left( \frac{x}{2 \xi} \right)^{1/4} 
{\rm e}^{\pi \xi -2 \sqrt{2 \xi x}}
+ \frac{i}{2} \left( \frac{x}{2 \xi} \right)^{1/4} 
{\rm e}^{2 \sqrt{2 \xi x} -  \xi \pi} 
\right]  \;\;,\;\; x \equiv - k \, \tau \;\;, \nonumber\\ 
&& {\bar A}_2  \left( \tau  ,\, k \right) = {\bar A}_1^*  \left( \tau  ,\, k \right) \;. 
\label{Ab12-sol} 
\end{eqnarray} 
These approximate solutions are normalized according to ${\bar A}_1 \, {\bar A}_2' - A_2 \, {\bar A}_1' = i$ (we note that this relation is satisfied exactly by (\ref{Ab12-sol})). This is due to the fact that the corresponding exact solutions satisfy this relation in the deep UV regime, and the quantity ${\bar A}_1 \, {\bar A}_2' - A_2 \, {\bar A}_1' $ is an integral of motion (namely, it is preserved by the second of (\ref{master})). An arbitrary phase in both of eqs.~(\ref{Ab12-sol}) has been fixed so that the amplified part of the solutions (the one that is exponentially large in the $\xi \gg 1$ limit) is real. The large real part was given in  \cite{Anber:2009ua} where it was shown how this approximation is obtained from the exact solution. The small imaginary part was given in \cite{Peloso:2016gqs}. In Appendix \ref{app:WKB} we show how the approximate solution (\ref{Ab12-sol}) emerges from a WKB approximation of the second of eqs.~(\ref{master}). 

For the purpose of computing the background solution of \cite{Anber:2009ua} only the exponentially large part of eq.~(\ref{Ab12-sol}) is needed, which coincides for both terms. Inserting ${\rm Re }
\left( {\bar A}_1 \right) = {\rm Re } \left( {\bar A}_2 \right)$ into the first of ({\ref{master}) and performing the integral\footnote{The integral is performed also in the $x \geq 2 \xi$ range where the expression (\ref{Ab12-sol}) is invalid. However, the integrand is strongly dominated by the peak of (\ref{Ab12-sol}), taking place in the $x \ll 2 \xi$ region where the solution is valid. Therefore, using the real part of  (\ref{Ab12-sol}) in the full domain results in an exponentially small mistake. More precisely, the integrand needs to be UV-regulated at large $x$, where the gauge mode is still in the vacuum mode, and using the real part (\ref{Ab12-sol}) also in this range effectively performs the regularization. A renormalization scheme for this model was provided in \cite{Ballardini:2019rqh}, with results in agreement with those obtained in this way. This happens because the physical peak from gauge field amplification is well separated from the vacuum, UV region.} results in~\cite{Anber:2009ua} 
\begin{equation} 
{\bar \Phi}'' + 2 a H \, {\bar \Phi}'  + a^2 \, V'  \simeq - \frac{a^2 \alpha}{f} \, 
\frac{315 \, H^4}{2^{17} \, \pi^2} \, \frac{{\rm e}^{2 \pi \xi}}{\xi^4} \;. 
\label{bck-eq}
\end{equation} 

In standard slow roll inflation the right hand side of this relation is not present, and $\ddot{\Phi}$ (where dot denotes derivative with respect to physical time) is negligible, so that the derivative of the inflaton potential is ``compensated'' by the Hubble friction term, $3 H \dot{\Phi} \simeq - V'$. In the AS solution the dominant friction is provided by the gauge field amplification, so that 
\begin{equation} 
\frac{315}{2^{17} \, \pi^2} \, \frac{\alpha H^4}{f} \, \frac{{\rm e}^{2 \pi \xi}}{\xi^4} \simeq - V' \;. 
\label{AS} 
\end{equation} 

In addition to this, we require that the energy density in the produced fields is much smaller than the inflaton potential energy, so to have an inflationary background. Using the mode functions (\ref{Ab12-sol}), disregarding the small imaginary part, this amounts to~\cite{Anber:2009ua} 
\begin{eqnarray} 
\frac{\rho_A }{V} = 
\frac{1}{4 \pi^2 a^4 \, V} 
\int d k \, k^2 \left[ \left\vert {\bar A}' \right\vert^2 + k^2  \left\vert {\bar A} \right\vert^2 \right]  
\simeq \frac{2}{5} \frac{f \xi}{\alpha} \frac{\left( - V' \right)}{V}  \ll 1 \;. 
\label{rhoAV}
\end{eqnarray} 

\section{Linearized system of perturbations} 
\label{sec:perturbations} 

We now study analytically small departures from the AS solutions of the system~(\ref{master}). To this goal, we decompose the inflaton and the gauge modes into the AS ones (those obtained in the previous section) plus small perturbations,  
\begin{equation}
\Phi = {\bar \Phi} + \delta \Phi \;\;\;,\;\;\; A = {\bar A} + \delta A \;\;,  
\label{deco}
\end{equation} 
and we solve the system  (\ref{master}) to first order in $\delta \Phi$ and $\delta A$. This procedure is not a complete perturbative study on the stability of the AS solution, since we disregard metric perturbations and spatial inhomogeneities of the inflaton. Nonetheless it captures the cases studied in the works \cite{Cheng:2015oqa,Notari:2016npn,DallAgata:2019yrr,Domcke:2020zez,Gorbar:2021rlt}, where the stability was studied numerically also assuming a homogeneous inflaton and no metric perturbations. It is hard to imagine that the inclusion of these two ingredients can make the AS solution stable, if an instability will emerge from the present analysis. In fact, the instability observed in~\cite{Cheng:2015oqa,Notari:2016npn,DallAgata:2019yrr,Domcke:2020zez,Gorbar:2021rlt} persists also in the lattice analysis of~\cite{Caravano:2022epk} which does include spatial fluctuations in the inflaton.

At first order in the perturbations (\ref{deco}), the system (\ref{master}) reads 
\begin{eqnarray} 
&& \delta \Phi'' + 2 a H \delta \Phi' + a^2 V'' \delta \Phi = - \frac{\alpha}{f a^2} \int \frac{d^3 k}{\left( 2 \pi \right)^3} \frac{k}{2} \, \frac{\partial}{\partial \tau} \left[ {\bar A} \, \delta A^* + {\bar A}^* \, \delta A \right] \;, \nonumber\\ 
&& \delta A'' + \left( k^2  - \frac{k \, {\bar \Phi}'}{f} \right) \delta A =  \frac{\alpha \, {\bar A}}{f} \, \delta \Phi' \;, 
\label{master-lin}
\end{eqnarray} 
where we note that we are also disregarding perturbations of $H$. As done in the previous section, we first formally solve the second equation for the gauge field modes as a functional of the inflaton derivative. This can be done via the Green function method, resulting in 
\begin{equation} 
\delta A \left( \tau ,\, k \right) = \frac{\alpha \, k}{f} \int^\tau d \tau'\, G_k \left( \tau ,\, \tau' \right) \, {\bar A} \left( \tau' ,\, k \right) \delta \Phi' \left( \tau' \right) \;, 
\label{dA-formal}
\end{equation} 
where the  Green function $G_k \left( \tau ,\, \tau' \right)$ is introduced and computed in Subsection 
\ref{subsec:green}. We then insert this formal solution into the first of eqs.~(\ref{master-lin}), that in this way becomes an integro-differential equation for the inflaton and its time derivatives 
\begin{align} 
\!\!\!\!\!\!\!\! 
\delta \Phi'' + 2 a H \delta \Phi' + a^2 V'' \delta \Phi =& - \frac{\alpha^2}{f^2 a^2} 
\int^\tau   d \tau'   \delta \Phi \left( \tau' \right) \nonumber\\
&
\times\frac{\partial}{\partial \tau} 
\int \frac{d^3 k}{\left( 2 \pi \right)^3} \frac{k^2}{2} \,  \left[ 
G_k^* \left( \tau ,\, \tau' \right) {\bar A} \left( \tau \right)  {\bar A}^* \left( \tau' \right) + {\rm c.c.}  \right] \;.
\label{eq-dPhi}
\end{align} 

In Subsection~\ref{subsec:source+sol}  we then work out the source term, and we obtain approximate analytical solutions for this  equation. 

\subsection{Green function} 
\label{subsec:green} 

Eq.~(\ref{dA-formal}) follows from eq.~(\ref{master-lin}), with the Green function satisfying  
\begin{equation} 
\left[ \frac{\partial^2}{\partial \tau^2} + k^2 - \frac{\alpha \, k \, \Phi'}{f} \right] G_k \left( \tau ,\, \tau' \right) = \delta \left( \tau - \tau' \right) \;, 
\label{eq-for-G}
\end{equation} 
where $\delta$ denotes the Dirac $\delta-$function. If ${\bar A}_{1,2}$ are two solutions of the associated homogeneous equation, it is immediate to see that the combination 
\begin{equation}
G_k \left( \tau ,\, \tau' \right) = \frac{{\bar A}_1 \left( \tau \right) {\bar A}_2 \left( \tau' \right) - 
{\bar A}_1 \left( \tau' \right) {\bar A}_2 \left( \tau \right)}{ {\bar A}_1' \left( \tau' \right) {\bar A}_2 \left( \tau' \right) - {\bar A}_1 \left( \tau' \right) {\bar A}_2' \left( \tau' \right)} \, \theta \left( \tau - \tau' \right) \;, 
\end{equation}
where $\theta$ is the Heaviside $\theta-$function, satisfies eq.~(\ref{eq-for-G}). This combination is the retarded Green function and it guarantees causality, since the Heaviside $\theta$-function ensures that only sources at times $\tau' < \tau$ can affect the solution at the time $\tau$. 

In the explicit construction of the Green function we use the two solutions of the homogeneous equation that reduce to (\ref{Ab12-sol}) in the $x \ll 2 \xi$ limit. As we already commented after those expressions, ${\bar A}_1 \, {\bar A}_2' - A_2 \, {\bar A}_1' = i$ for these solutions (this is valid at all $x$). We thus find 
\begin{eqnarray} 
G_k \left( \tau ,\, \tau' \right) 
&=& -\frac{\left( x \, x' \right)^{1/4}}{\sqrt{2 \xi} \, k} \, 
\sinh \left[ 2 \sqrt{2 \xi} \left( \sqrt{x} - \sqrt{x'} \right) \right] \theta \left( x' - x \right) 
\;\;\;,\;\;\; x,\, x' \ll 2 \xi \;, 
\label{Gsol-par}
\end{eqnarray} 
where we recall that $x \equiv - k \tau$ and $x' \equiv - k \tau'$. We note that, to obtain this expression, also the subleading terms in (\ref{Ab12-sol}) need to be retained (which is why we evaluated them, since only the dominant term is required for the AS background solution). 

We recall that the amplification of the gauge modes takes place deep inside the horizon. Therefore, as this region was regulated away from the integral in (\ref{master}) in the AS background solution, the same needs to be done in the present computation. We therefore multiply the expression (\ref{Gsol-par}) by two functions that vanish in the $x \gg 2 \xi$ and $x' \gg 2 \xi$ limits. We choose 
\begin{eqnarray} 
G_k \left( \tau ,\, \tau' \right) 
&\simeq& -\frac{\left( x \, x' \right)^{1/4}}{\sqrt{2 \xi} \, k} \, 
\sinh \left[ 2 \sqrt{2 \xi} \left( \sqrt{x} - \sqrt{x'} \right) \right] \theta \left( x' - x \right) \theta \left( 2 \xi \gamma^2 - x' \right) \theta \left( 2 \xi \gamma^2 - x \right)  \;\;, \nonumber\\
\label{Gsol-par2}
\end{eqnarray} 
where $\gamma$ is an order one constant that we keep unspecified, as a measure of the uncertainty associated with our regularization. We will see that our quantitative results are only weakly sensitive to $\gamma$, while our qualitative conclusions are insensitive to it. We note that the last $\theta-$function in this expression is superfluous, and therefore we write 
\begin{eqnarray} 
G_k \left( \tau ,\, \tau' \right) 
&\simeq& -\frac{\left( x \, x' \right)^{1/4}}{\sqrt{2 \xi} \, k} \, 
\sinh \left[ 2 \sqrt{2 \xi} \left( \sqrt{x} - \sqrt{x'} \right) \right] \theta \left( x' - x \right) \theta \left( 2 \xi\gamma^2 - x' \right) \;\;. 
\label{Gsol}
\end{eqnarray} 

\subsection{Linearized solutions} 
\label{subsec:source+sol}

We insert the expressions (\ref{Gsol}) and the dominant term of eq.~(\ref{Ab12-sol}) into the integro-differential equation~(\ref{eq-dPhi}).  Changing integration variables $ k \to y \equiv - 2 \xi k \tau'$, we obtain 
\begin{eqnarray} 
&& \delta \Phi'' + 2 a H \delta \Phi' + a^2 V'' \delta \Phi \nonumber\\ 
&& \quad\quad =  \frac{\alpha^2}{f^2 a^2} \frac{{\rm e}^{2 \pi \xi} }{2^8 \pi^2 \xi^5} 
\int^\tau  \frac{ d \tau' }{\left( - \tau' \right)^4}  \delta \Phi' \left( \tau' \right)  \,
\frac{\partial}{\partial \tau}  \int_0^{4 \xi_\gamma^2}  d y \, y^3 \,  \sqrt{\tau \tau'} 
\left[ {\rm e}^{-4 \sqrt{y}} - 
{\rm e}^{-4 \sqrt{y} \sqrt{\frac{-\tau}{-\tau'}}} \right] \;,\nonumber\\
\label{master-intdiff}
\end{eqnarray} 
where we have defined $\xi_\gamma\equiv \xi\gamma$. We look for ``power law'' solutions of the type 
\begin{equation}
\delta \Phi = C \, \left( - \tau \right)^{-\frac{1+\zeta}{2}} \;, 
\label{ansatz} 
\end{equation} 
where $C$ and $\zeta$ are constant. We expect that a solution of this type is possible since all terms in (\ref{master-intdiff}), in our working assumption of disregarding slow roll variations, evolve in time as $\delta \Phi / \tau^2$. 

As the system is linear, the constant $C \neq 0$ drops from the following analysis and it is irrelevant; we choose to denote the exponent with the combination $-\frac{1+\zeta}{2}$, rather than with a single symbol, as it simplifies some of the following algebra. We note that the result of this analysis will indicate that 
\begin{equation} 
{\rm AS \; solution \; is \; stable} \;\; \Leftrightarrow \;\; {\rm Re} \, \zeta < - 1 \;. 
\end{equation} 
Moreover, we need to impose ${\rm Re } \, \zeta > -8$, or the integral in eq.~(\ref{master-intdiff}) would diverge at $\tau' = - \infty$. Finally, we note that $\zeta$ might have an imaginary part. In fact, since eq.~(\ref{master-intdiff}) has real coefficients, a complex solution is always accompanied by its conjugate. Linear combinations of these solutions are therefore of the form (indicating explicitly the dependence on the Hubble rate to have a manifest dimensional consistency) 
\begin{equation}\label{inst_time_dep}
\delta \Phi \propto \left( - H  \tau \right)^{-\frac{1+ {\rm Re} \, \zeta}{2}} \, \cos \left( \frac{{\rm Im} \, \zeta}{2} \, \ln \left( - H  \tau \right) + \phi \right) \;, 
\end{equation} 
where $\phi$ is an arbitrary phase. Therefore a complex $\zeta$ corresponds to a solution that oscillates about the AS one, while converging to (for ${\rm Re} \, \zeta < - 1$) or departing from (for ${\rm Re} \, \zeta > - 1$) it. 

As we show in Appendix \ref{app:source}, inserting eq.~(\ref{ansatz}) into eq.~(\ref{master-intdiff}), performing the two integrals, and eliminating the common time dependence, results in 
\begin{eqnarray} 
&& \frac{\left( 1 + \zeta \right) \left( 7 + \zeta \right)}{4}   + \frac{ V'' }{H^2} \simeq   
\frac{\alpha \, \left( - V' \right)}{f \xi H^2} \, \frac{\left( 1 + \zeta \right) \left( 7 + \zeta \right)}{\zeta \left( 8 + \zeta \right)}  \left[  \frac{1}{\left( 8 \, \xi_\gamma \right)^\zeta} \, \frac{\Gamma \left( 9 + \zeta  \right)}{\Gamma \left( 9 \right)}  - 1  \right]  \;. 
\label{lineq-not}
\end{eqnarray} 

The first term in this relation originates from $\delta \Phi'' + 2 a H \delta \Phi' $. For standard slow roll inflation, $\left\vert V'' \right\vert \ll H^2$ and the right hand side vanishes. This results in $\zeta\simeq -1 ,\, -7$, namely $\delta  \Phi \propto \tau^0,\,\tau^2$, indicating the stability of the inflationary background. On the other hand, in the AS regime, $\left\vert V'' \right\vert \gg H^2$, so that the inflaton field is too massive to sustain inflation in absence of gauge field amplification. As we shall see, $\left\vert \zeta \right\vert = {\rm O } \left( 1 \right)$ also in this case. Therefore, for our study of stability, the first term in (\ref{lineq-not}) can be neglected, precisely as for the AS background (\ref{bck-eq}), leading to 
\begin{equation} 
\frac{\xi \, f \, V''}{\alpha \left( - V' \right)} \simeq  \frac{\left( 1 + \zeta \right) \left( 7 + \zeta \right)}{\zeta \left( 8 + \zeta \right)}  \left[  \frac{1}{\left( 8 \, \xi_\gamma \right)^\zeta} \, \frac{\Gamma \left( 9 + \zeta  \right)}{\Gamma \left( 9 \right)}  - 1  \right]  \equiv {\cal F} \left[ \zeta ,\, \xi_\gamma \right] \;. 
\label{calF}
\end{equation} 

The left hand side is a real quantity, whose sign depends on the sign of $V''$. Its magnitude is expected to be 

\begin{equation} 
\left\vert \frac{\xi \, f \, V''}{\alpha \left( - V' \right)} \right\vert = \left\vert \frac{V \, V''}{V^{' 2}} \times 
\frac{\xi \, f \left( - V' \right)}{\alpha \, V} \right\vert \ll 1 \;, 
\end{equation} 
thanks to the fact that the first factor is generically of order one, while the second factor is much smaller than one due to the condition (\ref{rhoAV}). 

Therefore we are looking for values of $\zeta$ for which the function ${\cal F} \left[ \zeta ,\, \xi_\gamma \right]$ is real, with a small absolute value. Let us start by studying the case of a real $\zeta$. We find the behavior shown in Figure~\ref{fig:calF-zetaR}, where ${\cal F}$ is plotted as a function of $\zeta$ for some representative values of $\xi_\gamma$. 

\begin{figure}[ht!]
\centerline{
\includegraphics[width=0.5\textwidth,angle=0]{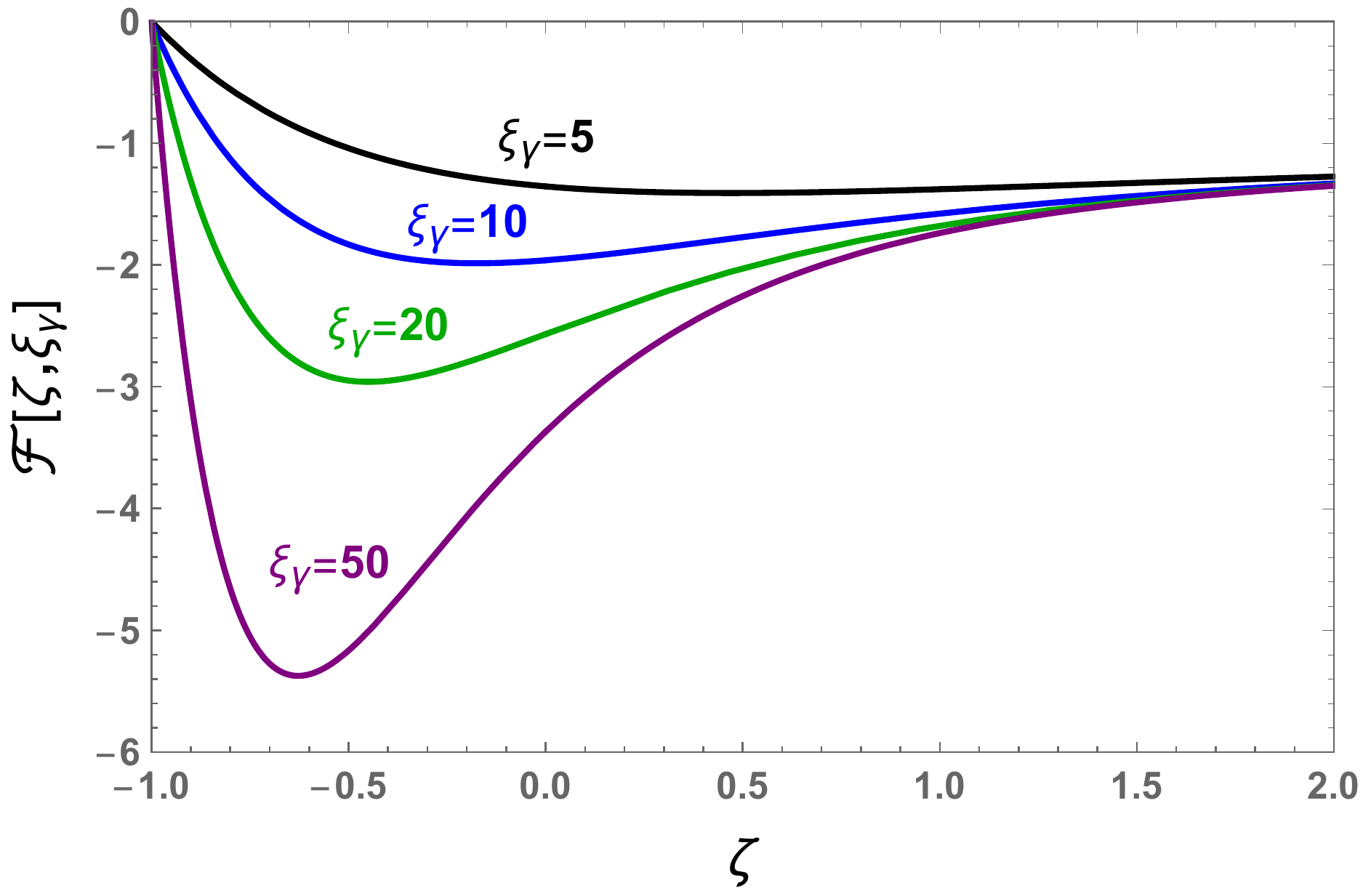}
}
\caption{Function ${\cal F}$, defined in eq. (\ref{calF}), as a function of a real parameter $\zeta$, for various values of $\xi_\gamma \equiv \xi \times \gamma$. }
\label{fig:calF-zetaR}
\end{figure}

We see that all real values of $\zeta$ lead to a negative ${\cal F}$, and therefore they can be solutions only for $V'' < 0$. The physically relevant region $\left\vert {\cal F} \right\vert \ll 1$ is obtained only for $\zeta$ slightly greater than $-1$, but very close to it. Expanding eq. (\ref{calF}) in this regime results in 
\begin{equation}
\zeta \simeq -1 - \frac{7}{6} \, \frac{1}{\gamma - 1 / \xi} \,  \frac{f \, V''}{\alpha \, \left( - V \right)} 
\simeq -1^+   \;\;\;\;,\;\;\;\; {\rm for \; }  V'' < 0 \;. 
\end{equation} 
From (\ref{ansatz}) we see that this mode is unstable, but the instability is extremely mild. 

Next, we study the case of a complex $\zeta$. We now show that, for any real and small value of ${\cal F}$, we can find an unstable mode with complex $\zeta$. For this purpose we focus our attention to the ${\rm Re } \left( \zeta \right) > -1$ case. We start by fixing $\xi_\gamma$ to a specific value, and by finding numerically the points in the $\left\{ {\rm Re } \left( \zeta \right) ,\, {\rm Im } \left( \zeta \right) \right\}$ plane resulting into a real ${\cal F} \left[ \zeta ,\, \xi_\gamma \right]$. 
For ${\rm Re} \left( \zeta \right) = - 1$, we find a finite set of values of ${\rm Im} \left( \zeta \right)$ for which this happens. We then find that the number of these values progressively decreases as 
${\rm Re} \left( \zeta \right)$ is increased, until a real ${\cal F}$ can no longer be found. The values of $\zeta$ for which ${\cal F}$ is real therefore form a set of distinct trajectories in the complex $\zeta-$plane. In Figure \ref{fig:zeRI} we show these trajectories for three specific choices of $\xi_\gamma$. We repeated this study for several values of $\xi_\gamma$ between those shown in the figure, always obtaining the same behavior as the one shown in the figure. 

\begin{figure}[ht!]
\centerline{
\includegraphics[width=0.33\textwidth,angle=0]{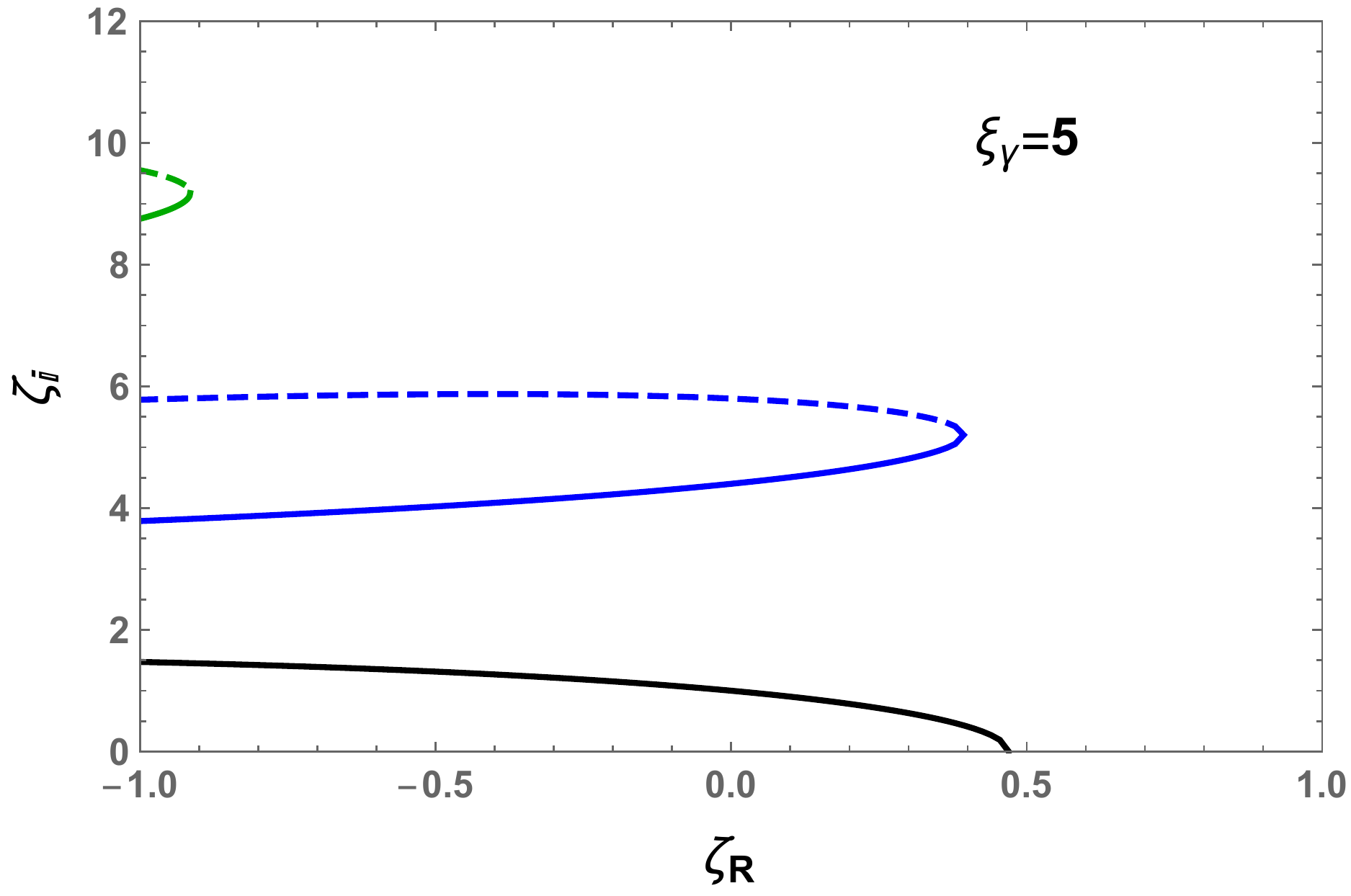}
\includegraphics[width=0.33\textwidth,angle=0]{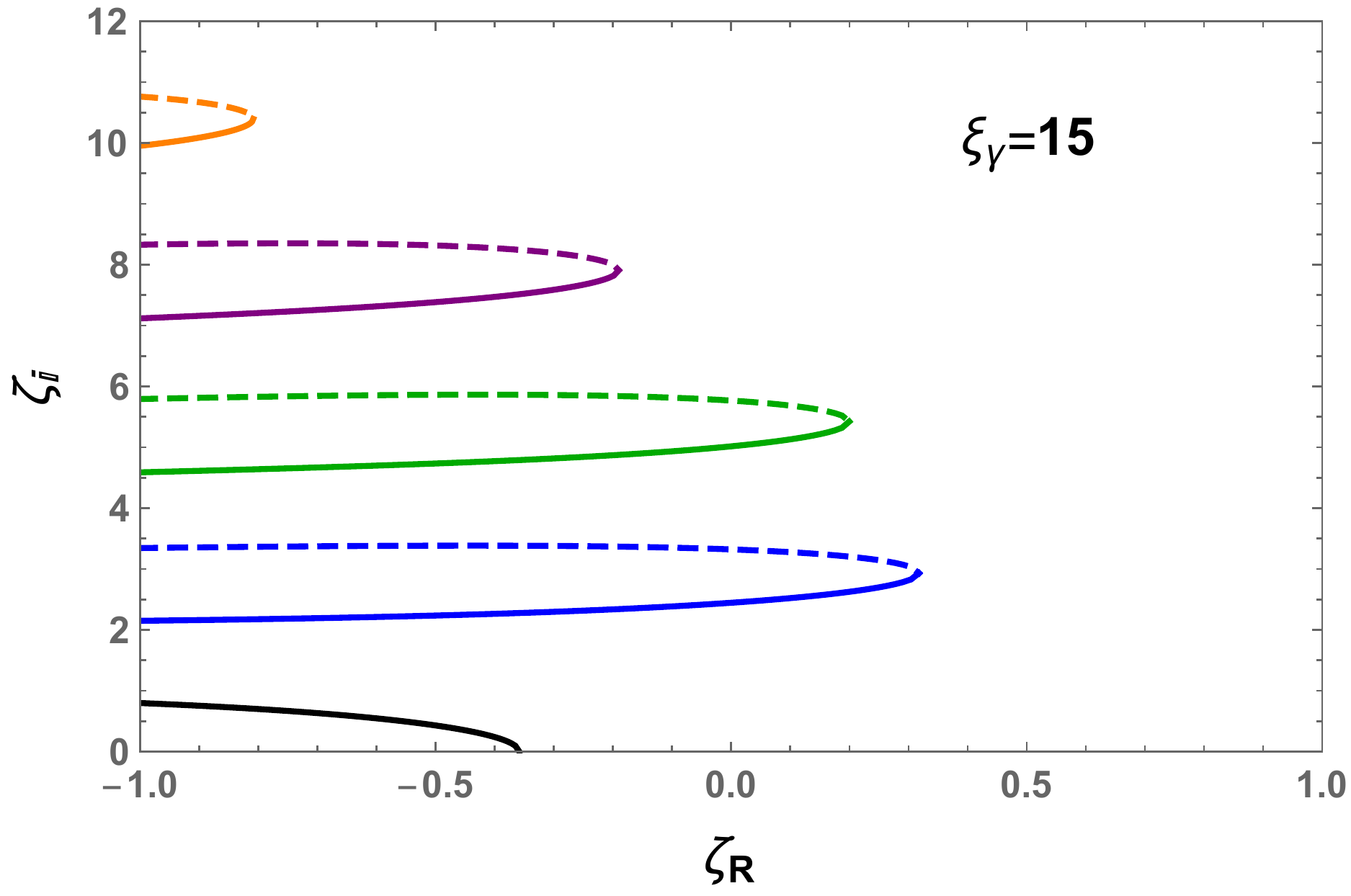}
\includegraphics[width=0.33\textwidth,angle=0]{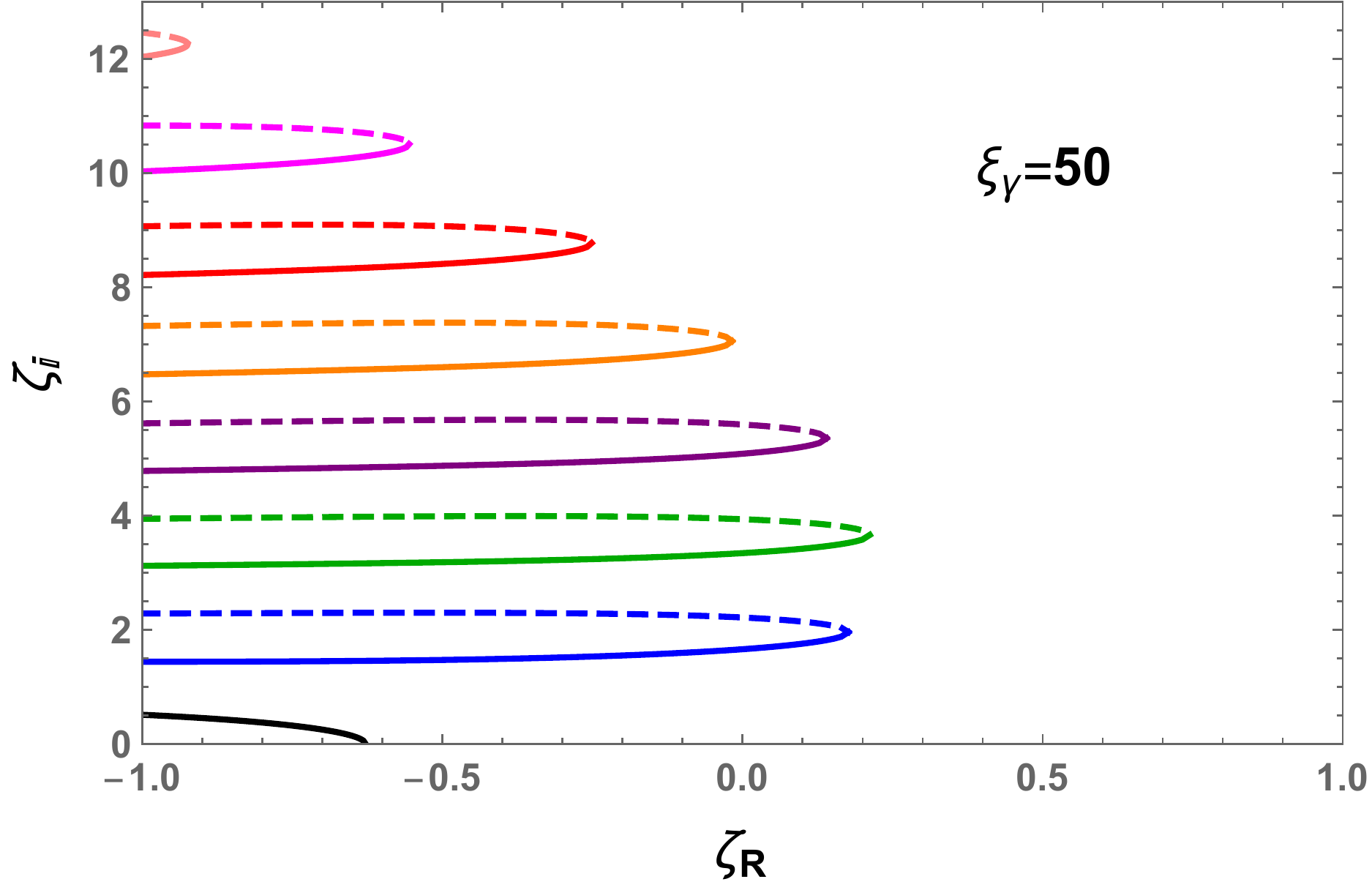}
}
\caption{Values of the real and imaginary parts of $\zeta$ resulting into a real ${\cal F} \left[ \zeta_R + i \, \zeta_I ,\, \xi_\gamma \right]$, for $\xi_\gamma = 5$ (left panel), $\xi_\gamma = 15$ (left panel),  and $\xi_\gamma = 50$ (right panel). These values form distinct trajectories in the complex $\zeta-$plane. The number of these trajectories increases as $\xi_\gamma$ increases. Beside the values shown in this figure, the function ${\cal F}$ is also real for ${\rm Im } \left( \zeta \right) = 0$ and for values of $\zeta$ that are the complex conjugate of those shown here. 
}
\label{fig:zeRI}
\end{figure}

Next, we evaluate ${\cal F}$ along the real-${\cal F}$ trajectories. 
In Figure \ref{fig:calF} we show ${\cal F}$ as a function of ${\rm Re } \left( \zeta \right)$ along the trajectories shown in the previous figure; namely, for each value of ${\rm Re } \left( \zeta \right)$, we first set ${\rm Im } \left( \zeta \right)$ according to one of the trajectories of Figure \ref{fig:zeRI} (coded with the same color and line style), and we then evaluate the corresponding value of ${\cal F}$. 

\begin{figure}[ht!]
\centerline{
\includegraphics[width=0.33\textwidth,angle=0]{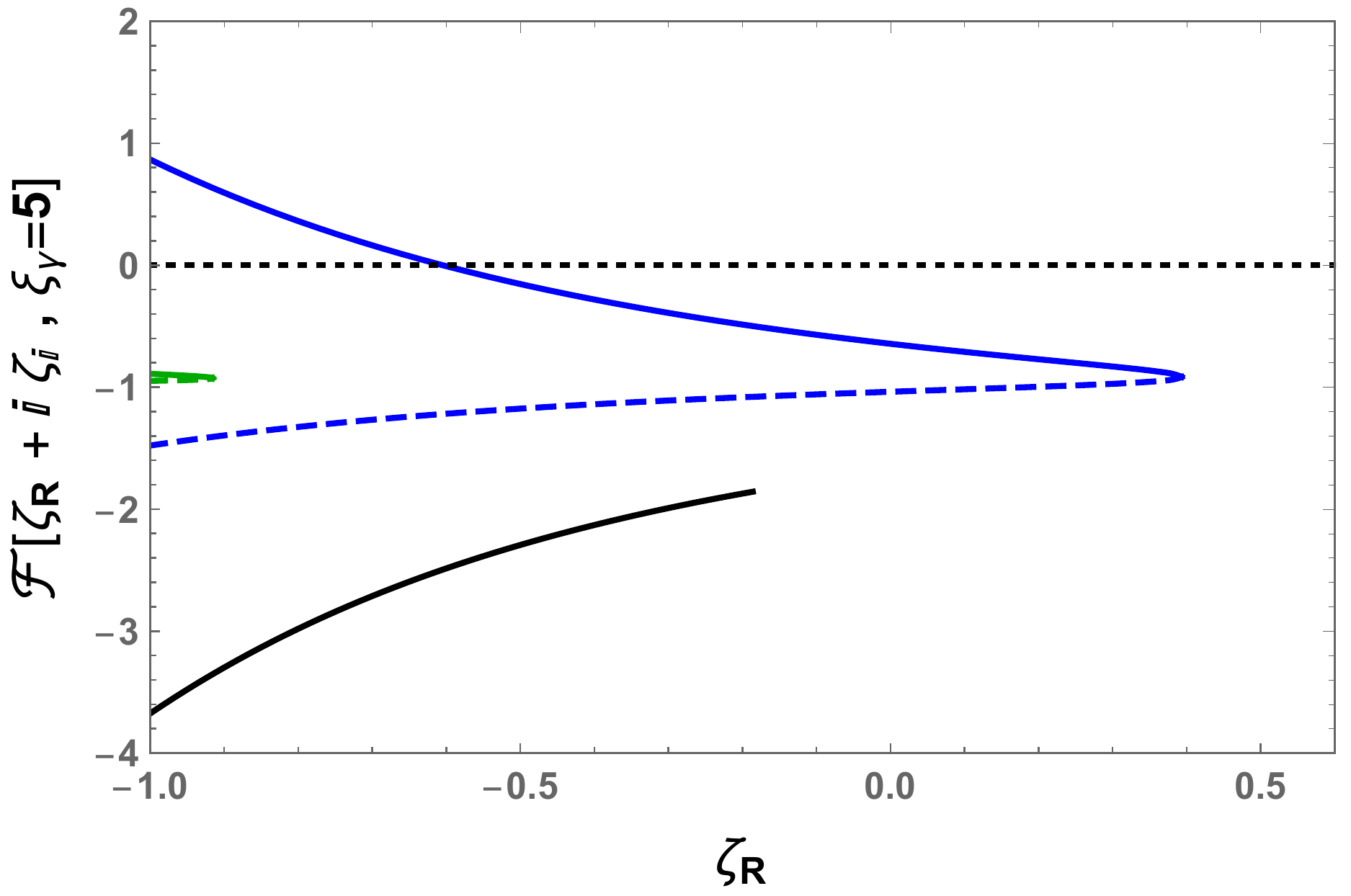}
\includegraphics[width=0.33\textwidth,angle=0]{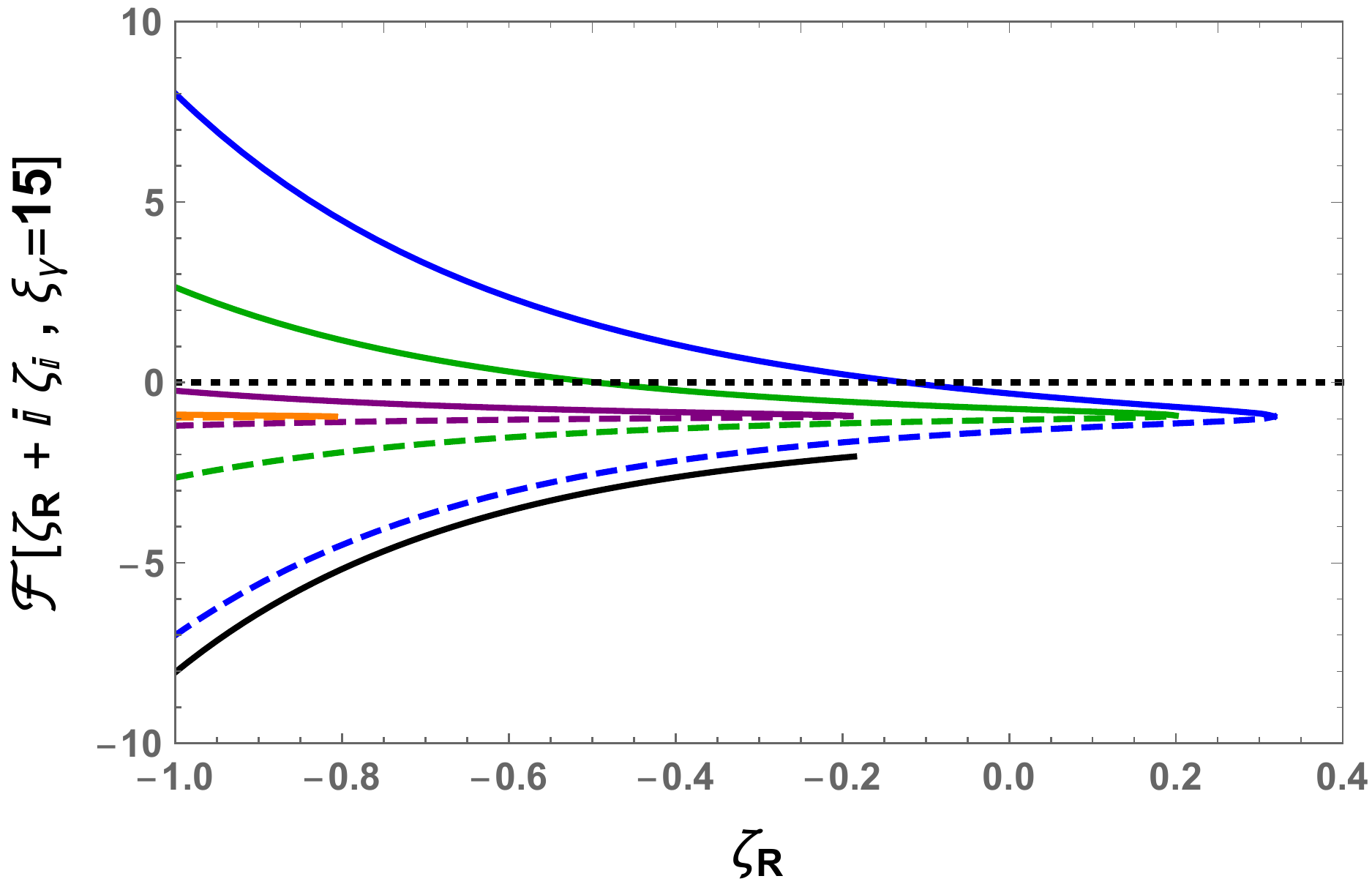}
\includegraphics[width=0.33\textwidth,angle=0]{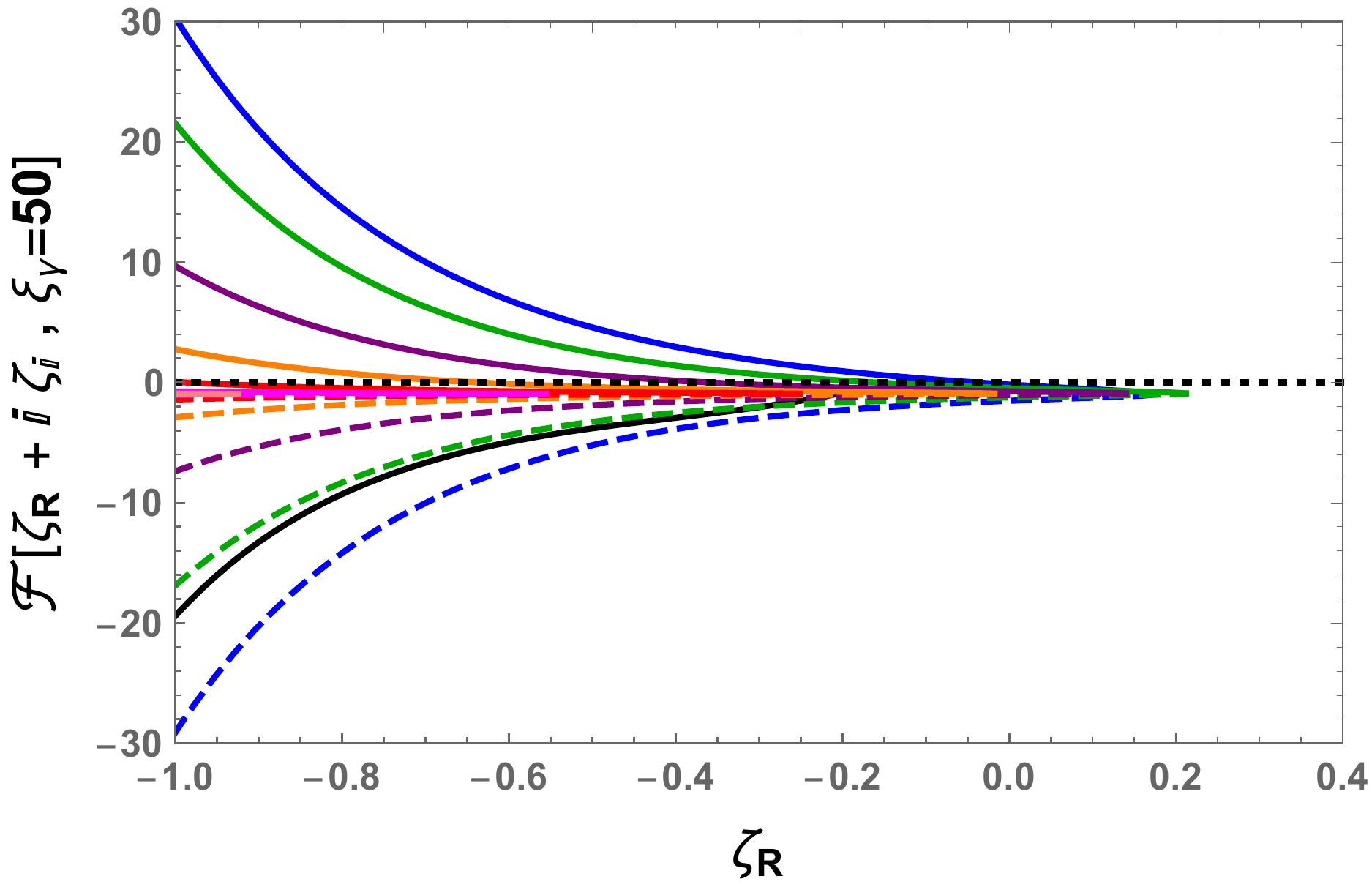}
}
\caption{Values of ${\cal F}$, shown as a function of ${\rm Re } \left( \zeta \right)$, along the trajectories shown in the previous figure. 
}
\label{fig:calF}
\end{figure}

We see from the figure that now a complex $\zeta$ solution is obtained also at positive $V''$. We recall that the physically relevant range is $\left\vert {\cal F} \right\vert \ll 1$. We also observe that, in the physically relevant range, ${\rm Re } \left( \zeta \right)$ is significantly greater than $-1$, resulting in a stronger instability than that obtained for real $\zeta$. A generic initial condition for $\delta \Phi$ will result in a linear combination of the solutions 
\begin{equation}
\delta \Phi = \sum c_i \, \left( - \tau \right)^{-\frac{1+\zeta_i}{2}} \;, 
\end{equation} 
where, in the general case, all the weights $c_i$ of the linear combination are nonvanishing. The instability will be then led by the $\zeta_i$ solution with the greatest real part.\footnote{We also note that all the solutions  we have found satisfy ${\rm Re } \, \zeta > -8$, and therefore the integral in eq. (\ref{master-intdiff}) is convergent.}

In the physical range $\left\vert {\cal F} \right\vert \ll 1$, the obtained values of $\zeta$ vary by a small amount. Therefore, the solution $\zeta$ depends mostly on the product $\xi_\gamma$, and only mildly on the $\frac{\xi \, f \, V''}{\alpha \left( - V' \right)}$ combination. As a consequence, we can obtain accurate solutions for $\zeta$ simply setting ${\cal F} \left[ \zeta ,\, \xi_\gamma \right] = 0$. In Figure \ref{fig:zetaR-zetaI} we then vary $\xi_\gamma$ and we show the solution for $\zeta$ (leading to ${\cal F} = 0$) with the largest real part, corresponding to the leading instability. 

\begin{figure}[ht!]
\centerline{
\includegraphics[width=0.5\textwidth,angle=0]{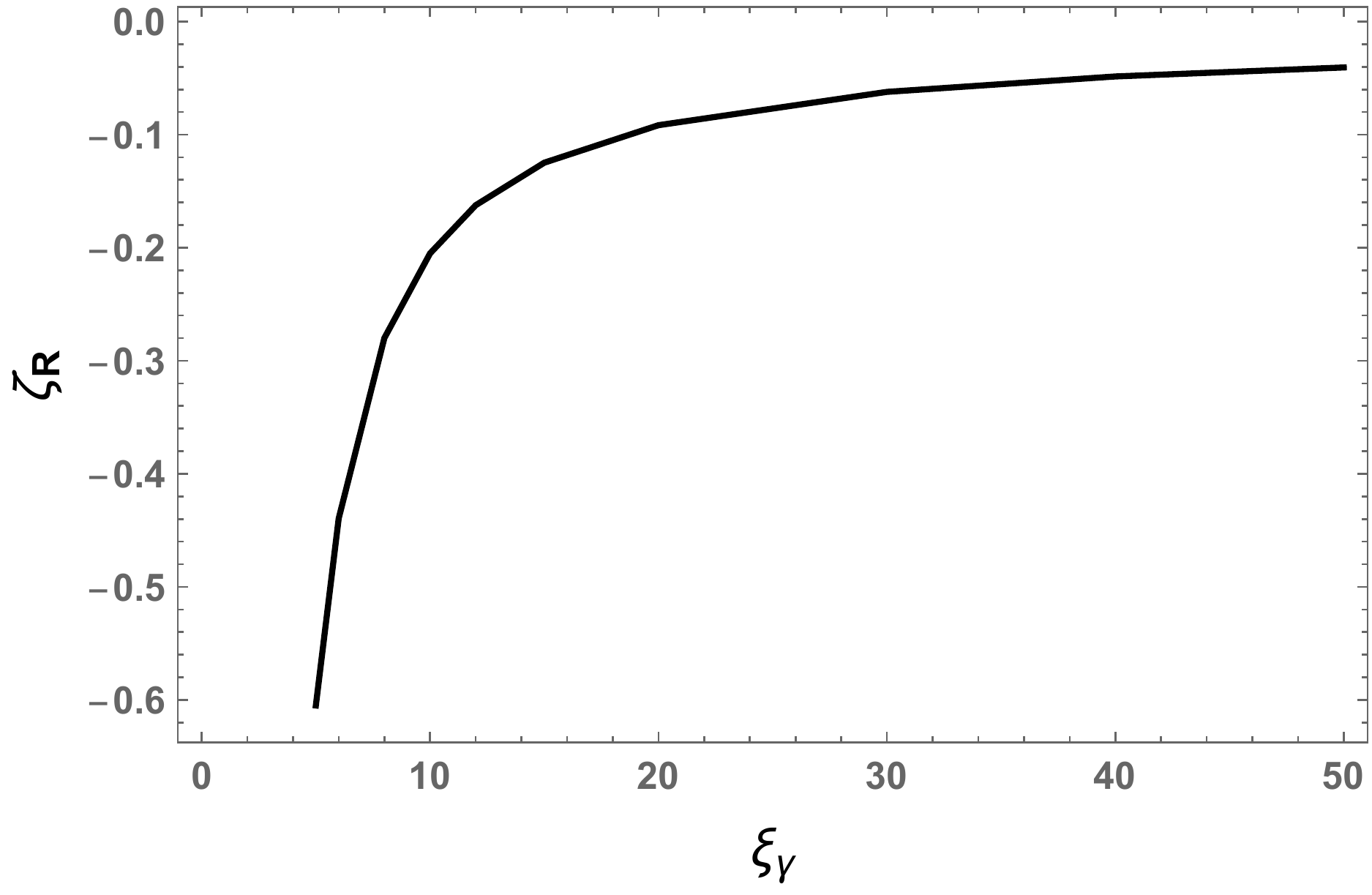}
\includegraphics[width=0.5\textwidth,angle=0]{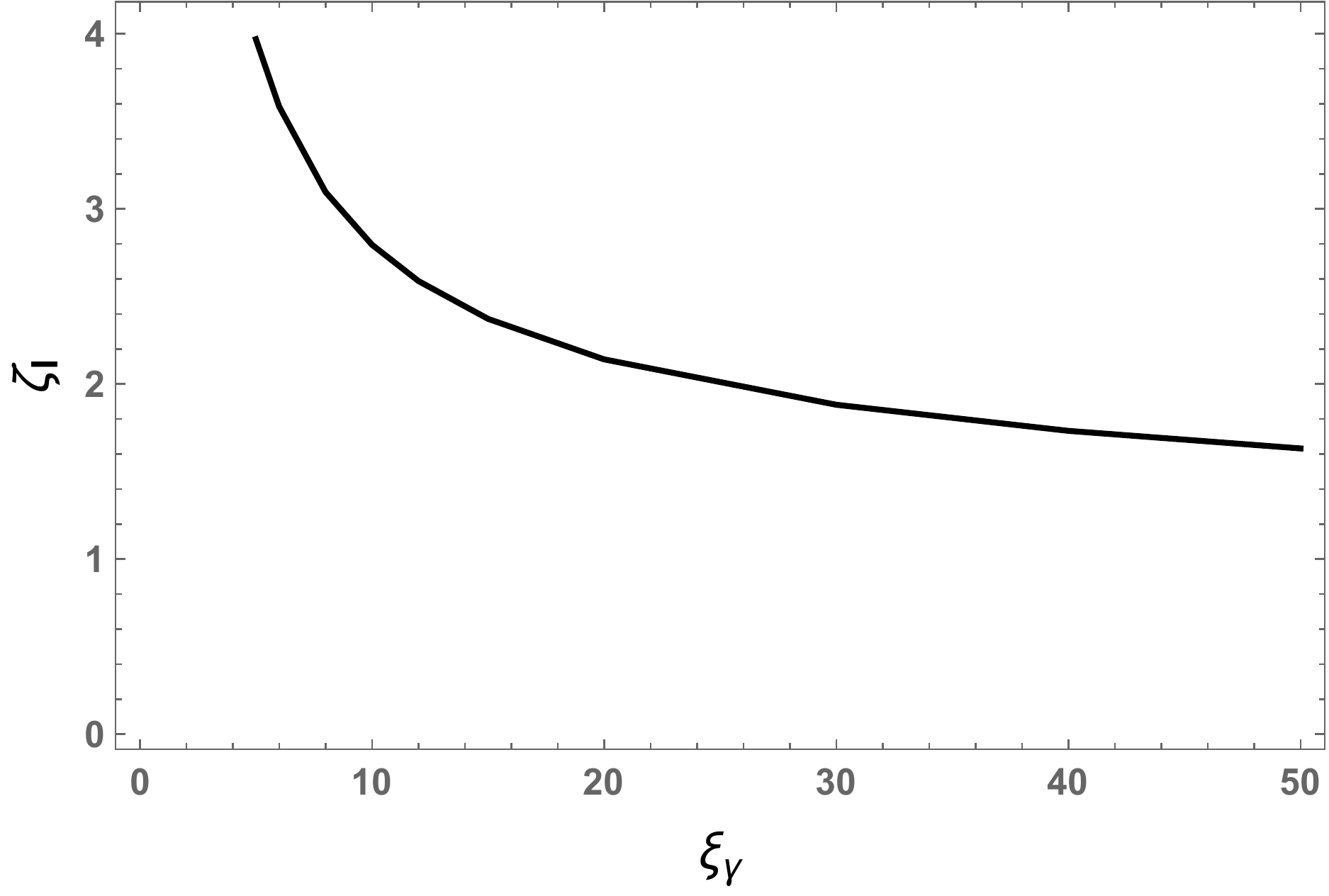}
}
\caption{Values of ${\rm Re } \left( \zeta \right)$ (left panel) and of ${\rm Im } \left( \zeta \right)$
(right panel) of the most unstable mode. 
}
\label{fig:zetaR-zetaI} 
\end{figure}

We recall that $\xi_\gamma = \gamma \, \xi$, where $\xi \equiv \frac{\alpha \, \dot{\bar \Phi}}{2 f  H}$ is the physically relevant parameter controlling the gauge field amplification, while $\gamma$ is a positive quantity of order one that measures the uncertainty of our computation (due to the regularization of the vacuum modes in the sub-horizon regime). We see that, not surprisingly, the instability becomes stronger at increasing $\xi$ (as it leads to a greater value of ${\rm Re } \left( \zeta \right)$, see eq.~(\ref{ansatz}) recalling that $\tau \to 0^-$ during inflation). We also see that our quantitative results are only mildly affected by the uncertainty encoded in $\gamma$. Most importantly, our conclusion about the instability of the AS background solution is unaffected by it. We have shown that the instability manifests itself as an oscillatory behavior about the AS solution with increasing amplitude, as also emerged from some previous numerical study \cite{Cheng:2015oqa,Notari:2016npn,DallAgata:2019yrr,Domcke:2020zez,Gorbar:2021rlt,Caravano:2022epk}.

\section{Conclusions} 
\label{sec:conclusions} 

We have presented the analytical study of the stability of the model of~\cite{Anber:2009ua} when one considers perturbations of the velocity of the zero mode of the inflaton about the mean field value determined by eq.~(\ref{AS}). Our results are consistent with previous findings in the literature, where the use of numerical techniques showed the existence of oscillations of increasing amplitude in the inflaton velocity when the system enters the strong backreaction regime. { Eq.~(\ref{inst_time_dep}), and the fact that Figure~\ref{fig:zetaR-zetaI} gives Im$\,\zeta\lesssim 2$, show that the oscillations have a period that is somehow less than $2\pi$ efoldings, which is in good agreement with the period of the oscillations of $\dot{\bar\Phi}$ found in the numerical works~\cite{DallAgata:2019yrr,Domcke:2020zez,Gorbar:2021rlt,Caravano:2022epk}.

Our work extends these numerical results, because we could show that the instability is present irrespective of the choice of parameters in the model, as long as one is in the strong backreaction regime. While we did not consider spatial fluctuations in the inflaton field, we do not expect these to change this picture - in fact, large oscillations in the inflaton velocity have also been observed in~\cite{Caravano:2022epk}, that studied the system on a lattice accounting also for the space dependence of the inflaton. 

As first observed in~\cite{Domcke:2020zez}, the origin of the instability can be traced to the fact that the source term on the right-hand side of the first of eqs.~(\ref{master}) has a delayed response to a change in the inflaton velocity. Our eq.~(\ref{master-intdiff}) makes the origin of this delay clear, as the integral in $dy$ in that equation is peaked at values of $\tau'$ different from $\tau$. We would not be surprised if this behavior is common to other models where the inflaton is slowed down by the backreaction of produced matter.

\vskip.25cm
\section*{Acknowledgements} 

M.P. is supported by Istituto Nazionale di Fisica Nucleare (INFN) through the Theoretical Astroparticle Physics (TAsP) and the Inflation, Dark Matter and the Large-Scale Structure of the Universe (InDark) project. The work of L.S. is partially supported by the US-NSF grant PHY-2112800.

\vskip.25cm

\appendix

\section{WKB approximate solutions for the background gauge field}
\label{app:WKB}

The second of eqs.~(\ref{master}) can be written in the form 
\begin{eqnarray}
\frac{d^2 {\bar A}}{d x^2} + \omega^2 \, {\bar A} = 0 \;\;,\;\; 
\omega^2 \equiv 1 - \frac{2 \xi}{x} \;,\; 
\label{eqAb-WKB}
\end{eqnarray}
where we recall that $x \equiv - k \tau$ formally ranges from $+\infty$ (the sub-horizon regime) to $0^+$ (the super-horizon regime). We note that $\omega^2$ is positive for $x > x_0  \equiv 2 \xi$ (in the WKB approximation of the Schr\"odinger equation, this corresponds to the ``classically allowed region'', where the solution is oscillatory), it vanishes at $x =x_0$, and is negative for $x < x_0$ (the ``classically forbidden region'', where the solution is the sum of two exponentials).  Apart from a neighborhood of $x_0$, the ``frequency'' varies adiabatically, $\left\vert \frac{1}{\omega^2} \, \frac{d \omega}{d x} \right\vert \ll 1$, and the equation can be solved approximately.  The WKB method relates the approximate solution in the ``allowed region'' to that in the ``forbidden region'' according to 
\begin{eqnarray}
{\bar A} &=& \frac{\alpha}{\left( \omega^2 \right)^{1/4}} \, \cos \left( \int_{x_0}^x \sqrt{\omega^2} \, d x '- \frac{\pi}{4} \right) -  \frac{\beta}{\left( \omega^2 \right)^{1/4}} \, \sin \left( \int_{x_0}^x \sqrt{\omega^2} \, d x' - \frac{\pi}{4} \right) \;\;,\;\; x \gg x_0 \;, \nonumber\\ 
{\bar A} &=& \frac{\alpha/2}{\left( -\omega^2 \right)^{1/4}} \, \exp \left( - \int_x^{x_0} \sqrt{-\omega^2} \, d x'  \right) +  \frac{\beta}{\left( - \omega^2 \right)^{1/4}} \, \exp \left( \int_x^{x_0} \sqrt{-\omega^2} \, d x' \right) \;\;,\;\; x \ll x_0 \;, \nonumber\\ 
\label{WKB}
\end{eqnarray} 
where $\alpha$ and $\beta$ are integration constants. 

We fix the integration constants by demanding the standard adiabatic mode in the deep sub-horizon regime 
\begin{equation}
\lim_{x \to + \infty} {\bar A} = i \, \frac{{\rm e}^{i \int_{2 \xi}^x \sqrt{1-\frac{2 \xi}{x'}} d x' - \frac{i \pi}{4}}}{\sqrt{2 k \sqrt{1-\frac{2 \xi}{x}}}} \;, 
\label{adiabatic} 
\end{equation} 
where an arbitrary phase has be chosen so that, once inserted in the first line of (\ref{WKB}), the growing mode in the corresponding second line is real. We note that the adiabatic mode (\ref{adiabatic}) is indeed of the form of the first line of (\ref{WKB}), with $- i \alpha =  \beta = \frac{1}{\sqrt{2 k}}$. We insert these coefficients into the second line of  (\ref{WKB}), and we perform the integration 
\begin{equation} 
\int_x^{2 \xi} \sqrt{\frac{2 \xi}{x'} - 1} \, d x' = 2 \xi \, \arctan \left( \sqrt{\frac{2 \xi}{x}  - 1 } \right) - \sqrt{2 \xi x - x^2} \;\;,\;\; 0 < x < 2 \xi \;. 
\end{equation} 
Moreover, in this regime, we can approximate 
\begin{equation} 
x \ll 2 \xi \;\; \Rightarrow \;\; {\rm arctan} \, \left( \sqrt{\frac{2 \xi}{x}-1} \right)  \simeq \frac{\pi}{2} - \sqrt{\frac{x}{2 \xi}} \;\;,\;\; \sqrt{2 \xi x-x^2} \simeq \sqrt{2 \xi x} \;. 
\end{equation} 

Considering all this, the second line of (\ref{WKB}) becomes 
\begin{equation} 
{\bar A} \simeq \frac{i}{\sqrt{2 k}} \frac{1}{2} \left( \frac{x}{2 \xi} \right)^{1/4} \, {\rm exp} \left[ - \pi \xi + 2 \sqrt{2 \xi x} \right]  +  \frac{1}{\sqrt{2 k}} \left( \frac{x}{2 \xi} \right)^{1/4} \, {\rm exp} \left[  \pi \xi - 2 \sqrt{2 \xi x} \right]  \;\;,\;\; x \ll 2 \xi \;, 
\label{sol-WKB}
\end{equation} 
which coincides with the expression ${\bar A}_1$ given in eq.~(\ref{Ab12-sol}) of the main text. We note that the second term in the expression~(\ref{sol-WKB}), which is the exponentially large term in the $\xi \gg 1$ regime, is indeed real.  Since the coefficient in eq.~(\ref{eqAb-WKB}) is real, also the complex conjugate of (\ref{sol-WKB}) is an approximate solution of this equation. This is the ${\bar A}_2$ solution given in the main text. We note that the two solutions are linearly independent.

\section{Evaluation of the source term}
\label{app:source}

In this Appendix we derive eq.~(\ref{lineq-not}) starting from eq.~(\ref{master-intdiff}), and, in particular, we evaluate the integrals in the source term. We start by inserting the ansatz (\ref{ansatz}) into eq.~(\ref{master-intdiff}), eliminating the common $C$ factor. The integral in the resulting expression is then evaluated as 
\begin{eqnarray} 
&& \int^\tau  d \tau'  \left( - \tau' \right)^{-5 - \frac{\zeta}{2}}  \,
\frac{\partial}{\partial \tau}  \int^{4 \xi_\gamma^2}  d y \, y^3 \, 
\left( - \tau \right)^{1/2} 
\left[ {\rm e}^{-4 \sqrt{y}} - 
{\rm e}^{-4 \sqrt{y} \sqrt{\frac{-\tau}{-\tau'}}} \right] = \frac{\left( - \tau \right)^{- \frac{9+\zeta}{2}}}{2^{12}} 
\Bigg\{ 
- \frac{4410}{\zeta} - \frac{630}{8 + \zeta} 
\nonumber\\ 
&& 
+ \frac{2 \, {\rm e}^{-8 \xi_\gamma}}{8 + \zeta} \left[ 315 + 2520 \xi_\gamma + 10080 \xi_\gamma^2 + 26880 \xi_\gamma^3 + 53760 \xi_\gamma^4 + 86016 \xi_\gamma^5 + 114688 \xi_\gamma^6 + 131072 \xi_\gamma^7  \right] \nonumber\\ 
&& + \frac{ \left( 7 + \zeta \right) \, \Gamma \left( 8 + \zeta \right)}{2^{3 \left( 1 + \zeta \right)} \, \zeta \,  \xi_\gamma^\zeta} - \frac{1}{2^{3 \left( 1 + \zeta \right)}  \xi_\gamma^\zeta } \Big[ 
35280 \, \Gamma \left[ \zeta , 8 \xi_\gamma \right] 
+ 17640 \, \Gamma \left[ 2 +  \zeta, 8 \xi_\gamma \right]  \nonumber\\ 
&& \quad\quad 
+ 5880 \, \Gamma \left[ 3 +  \zeta, 8 \xi_\gamma \right] 
+ 1470 \, \Gamma \left[ 4 +  \zeta, 8 \xi_\gamma \right] 
+ 294 \, \Gamma \left[ 5 +  \zeta, 8 \xi_\gamma \right] 
+ 49 \, \Gamma \left[ 6 +  \zeta, 8 \xi_\gamma \right]   \nonumber\\ 
&& \quad\quad 
+ 7 \, \Gamma \left[ 7 +  \zeta, 8 \xi_\gamma \right] 
+  \Gamma \left[ 8 +  \zeta, 8 \xi_\gamma \right] 
+  35280 \, \Gamma \left[  1 + \zeta , 8 \xi_\gamma \right] \Big] \Bigg\} \;, 
\end{eqnarray} 
where the functions $\Gamma$ with two arguments appearing in the last three lines are incomplete 
$\Gamma-$functions. This expression is valid for ${\rm Re } \, \zeta > -8$, where the integral converges. 

Disregarding the terms that are exponentially small in the limit of large $\xi_\gamma$  (including the incomplete $\Gamma$-functions) results in 
\begin{eqnarray} 
&& \int^\tau  d \tau'  \left( - \tau' \right)^{-5 - \frac{\zeta}{2}}  \,
\frac{\partial}{\partial \tau}  \int^{4 \xi_\gamma^2}  d y \, y^3 \, 
\left( - \tau \right)^{1/2} 
\left[ {\rm e}^{-4 \sqrt{y}} - 
{\rm e}^{-4 \sqrt{y} \sqrt{\frac{-\tau}{-\tau'}}} \right] \nonumber\\ 
&& \quad\quad\quad\quad \simeq 
\frac{\left( - \tau \right)^{- \frac{9+\zeta}{2}}}{2^{12}} 
\left\{   
- \frac{4410}{\zeta} - \frac{630}{8 + \zeta} 
+ \frac{\left( 7 + \zeta \right) \Gamma \left( 8 + \zeta \right)}{2^{3 \left( 1 + \zeta \right)} \, \zeta \, \xi_\gamma^\zeta } \right\} \;. 
\end{eqnarray} 

Inserting this expression into the right hand side of (\ref{master-intdiff}) results in 
\begin{eqnarray} 
&& \frac{\left( 1 + \zeta \right) \left( 7 + \zeta \right)}{4}  \left( - \tau \right)^{- \frac{5+\zeta}{2}}    + \frac{ V'' }{H^2 \tau^2} \left( - \tau \right)^{- \frac{5+\zeta}{2}}    \nonumber\\ 
&& =  \frac{\alpha^2 \left( - H \tau \right)^2}{f^2 } \frac{{\rm e}^{2 \pi \xi} }{2^8 \pi^2 \xi^5} 
\frac{\left( - \tau \right)^{- \frac{9+\zeta}{2}} }{2^{12}} \, 
\frac{\left( 1 + \zeta \right) \left( 7+\zeta \right)}{4 \zeta \left( 8 + \zeta \right)} \left[ -2^5 \times 315 + \frac{\Gamma \left( 9 + \zeta \right)}{ 2^{2+3 \zeta} \, \xi_\gamma^\zeta} \right] \;.
\end{eqnarray} 
Removing the common time dependence, and using eq.~(\ref{AS}) results into the expression~(\ref{lineq-not}) of the main text.

\end{document}